\documentclass[useAMS,usenatbib,usegraphicx]{mn2e}
\usepackage{epsfig}
\usepackage{amsmath} 
\usepackage{rotating}           
\usepackage{color}     
\usepackage{graphicx}
\usepackage{times}

\def\lapp{\ifmmode\stackrel{<}{_{\sim}}\else$\stackrel{<}{_{\sim}}$\fi}
\def\gapp{\ifmmode\stackrel{>}{_{\sim}}\else$\stackrel{>}{_{\sim}}$\fi}

\title[Quasar photometric redshifts from incomplete data]{Quasar photometric redshifts from incomplete data using Deep Learning}

\author[S. J. Curran]{S. J. Curran\thanks{Stephen.Curran@vuw.ac.nz}\\
School of Chemical and Physical Sciences, Victoria University of Wellington, PO Box 600, Wellington 6140, New Zealand}

\begin{document}

 \date{Accepted ---. Received ---; in original form ---}

\pagerange{\pageref{firstpage}--\pageref{lastpage}} \pubyear{2022}

\maketitle
\label{firstpage}
\begin{abstract}
Forthcoming astronomical surveys are expected to detect new sources in such large numbers that measuring
their spectroscopic redshift measurements will be not be practical. Thus, there is much interest in using machine
learning to yield the redshift from the photometry of each object.  We are particularly 
interested in radio sources (quasars) detected with the {\em Square  Kilometre Array}  and 
have found  Deep Learning, trained upon a large optically-selected sample of
 quasi-stellar objects, to be effective in the prediction of the redshifts in three external samples 
of radio-selected sources.  However, the requirement of nine different magnitudes, from the near-infrared, optical and ultra-violet bands,
 has the effect of significantly reducing the number of sources for which
  redshifts can be predicted. Here we explore the possibility of using machine learning to impute the missing features.
  We find that for the training sample,  simple imputation is sufficient, particularly replacing the missing
  magnitude with the maximum for that band, thus presuming that the non-detection is at the sensitivity limit. For the test
  samples, however, this does not perform as well as multivariate imputation, which suggests that many of the missing
  magnitudes are not limits, but have indeed not been observed.  From extensive testing of the models, we suggest that the
  imputation is best restricted to two missing values per source. Where the sources overlap on the sky, in the worst case,
  this increases the fraction of sources for which redshifts can be estimated from 46\% to 80\%, with $>90$\% being
  reached for the other samples. 
\end{abstract}  

\section{Introduction} 
\label{intro}

Given that large number of sources expected to be detected by forthcoming continuum surveys with the next generation of
large telescopes (e.g. \citealt{nha+11,asu+17,lnp18}), there is currently much interest in using machine learning
techniques to determine their redshifts from their photometry. 
These generally utilise the $u-g$, $g - r$, $r - i$ and $
i - z$ colours as features to train and validate upon sources in the {\em Sloan Digital Sky Survey} (SDSS,
e.g. \citealt{rws+01,wrs+04,mhp+12,hdzz16}). The addition of other bands, specifically the near-infrared (NIR) $W1$ and $W2$
bands and the ultra-violet (UV) $NUV$ and $FUV$ bands, can greatly improve the accuracy of the predictions, reducing the scatter (see Sect. \ref{dl}) between the
predicted and actual redshifts to $\sigma_{\Delta z ({\rm norm})}\sim0.1$ \citep{bmh+12,bcd+13,sih19,cmp21}.  
This is similar
to the scatter reached by state-of-the-art template fitting of the SEDs (\citealt{hac+10,bdb+17,bsf+20} and references therein), 
although the samples are generally smaller ($n\lapp10^3$, cf. $\gapp10^4$) and often require the prior removal of outliers.

From our previous results \citep{cm19,cur20}, we \citep{cmp21} suggested that the wide range of magnitudes was required
to cover the redshifting of rest-frame features, for example, the $\lambda\sim1~\mu$m inflection 
and the $\lambda= 1216$~\AA\ Lyman-break.  
This signifies a commonality between the data driven
(machine learning) methods and the physically motivated (template fitting) methods of photometric redshift
estimation \citep{sih19}.  

The requirement of a measurement in each of the nine bands has the effect of reducing the sample size. For example, of
the $100\,000$ strong training sample only $\approx70\%$ have the full complement of measurements in all of the
$FUV,NUV,u,g,r,i,z,W1,W2$ bands \citep{cmp21}. This worsens for the test data, which comprise three external samples of
radio selected surveys, chosen to test the potential of using an SDSS trained model to predict the redshifts of {\em Square  Kilometre Array} (SKA) 
(pathfinder) data (see Sect.~\ref{test_data}). The common practice is to remove sources for which all of the photometry
is not available\footnote{Exceptions are \citet{bmh+12}, who use the probability distribution in the feature--redshift
  space of the other sources to fill in the missing values, and \citet{cma+21}, who assume the detection limit of that
  band for the missing values.} , which can have the effect of dramatically reducing the sample size (e.g. up to half of
one of our training sets).  In this paper, we explore the possibility of using machine learning methods to mitigate the
effect of missing data, reducing the number of sources for which we cannot obtain photometric redshifts.

\section{Analysis}
\subsection{The training data}

\subsubsection{The data}
\label{sec:un}

For the training data  we extracted the first 100\,337 quasi-stellar objects (QSOs) with accurate spectroscopic redshifts ($\delta z/z<0.01$) 
from the SDSS Data Release 12 (DR12, \citealt{aaa+15}). We then matched the nearest source within 6 arc-seconds
in the {\em NASA/IPAC Extragalactic Database} (NED), usually resulting in a single match.   For these, the photometry was scraped 
from NED, WISE \citep{wem+10}, the {\em Two Micron All Sky Survey} (2MASS, \citealt{scs+06}) and GALEX \citep{bst17} databases. 
As per \citet{cur20,cmp21}, in order to ensure a uniform magnitude measure between the SDSS and other samples, 
for each QSO we added the PSF
flux densities associated with the AB magnitudes which fell within $\Delta\log_{10}\nu = \pm0.05$ of the central
frequency of the band.  Within each band range the fluxes were then averaged before being converted to a
magnitude.  This method, rather than just using the SDSS magnitudes directly, gives the option of using the SDSS data to
train other samples, for which direct SDSS photometry may not be available.

Extensive testing showed that use of all $FUV,NUV,u,g,r,i,z,W1,W2$ bands gave the most accurate photometric redshifts,
with the addition of the WISE $W3,W4$ photometry having little benefit, while significantly decreasing the sample size
\citep{cmp21}. However, even with the exclusion of the $W3,W4$ bands, there were only 72\,276 QSOs which had all of the
required photometry (Table~\ref{missing}).  
\setlength{\tabcolsep}{0.5em}
\begin{table}
\centering
\caption{The number of missing magnitude measurements in each of the bands of the 100\,337 strong SDSS QSO sample.  As expected, the numbers are larger for  the non-SDSS photometry.}
\begin{tabular}{@{}c c c c c c c c c c @{}}
\hline
\smallskip
 $FUV$ &    $NUV $  & $u$ &  $g$ &    $r$  &     $i$ &   $z$ &    $W1$ &    $W2$ &    \\
10\,658 & 20\,300 & 2923& 1301 & 1230 & 1256 & 1258 & 4040 & 4156\\
\hline
\end{tabular}
\label{missing}  
\end{table} 

\subsubsection{Data imputation}
\label{sec:methods}

\begin{figure*}
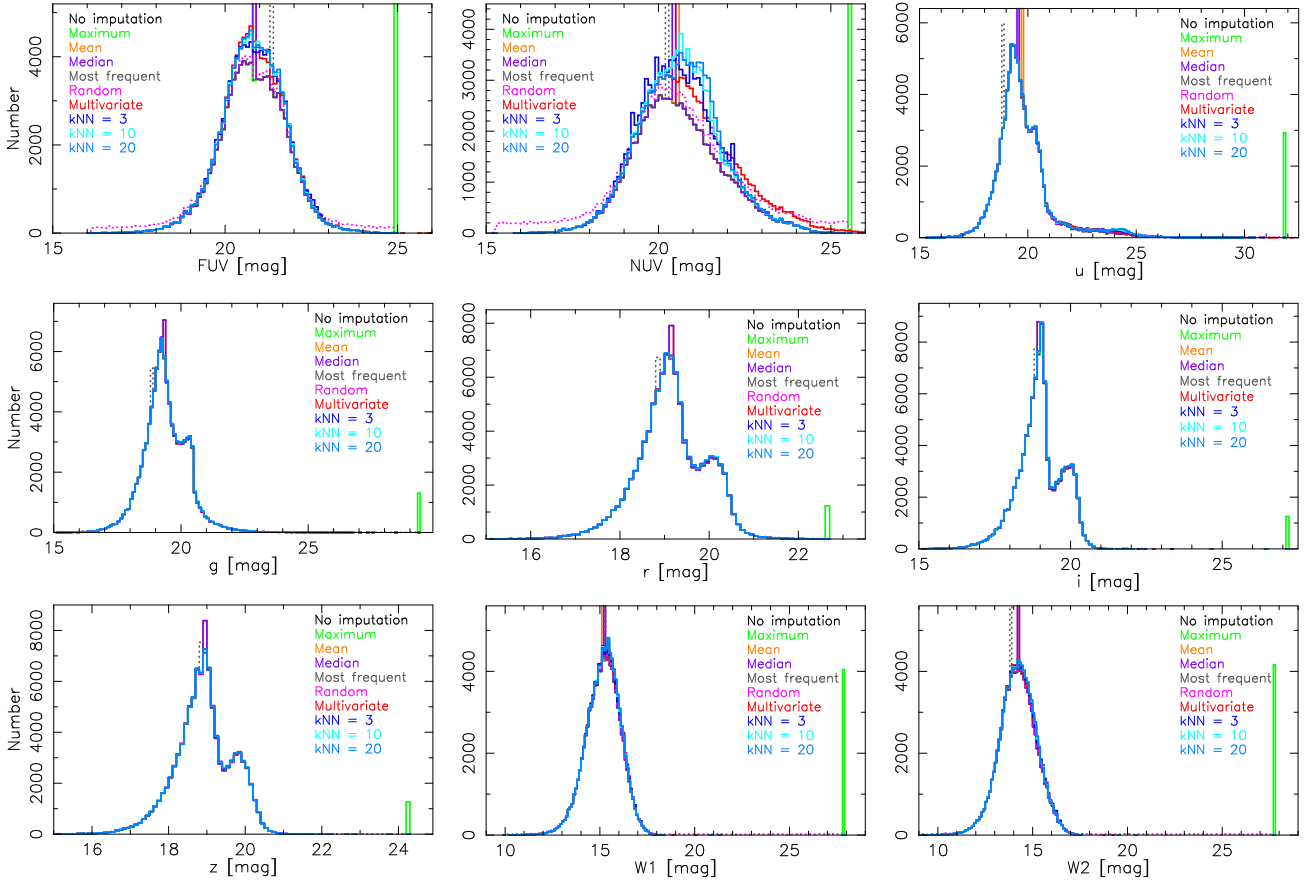

\centering \includegraphics[angle=-90,scale=0.35]{FUV-histo_110.eps}
\centering \includegraphics[angle=-90,scale=0.35]{NUV-histo_110.eps}
\centering \includegraphics[angle=-90,scale=0.35]{u-histo_175.eps}
\centering \includegraphics[angle=-90,scale=0.35]{g-histo_149.eps}
\centering \includegraphics[angle=-90,scale=0.35]{r-histo_85.eps}
\centering \includegraphics[angle=-90,scale=0.35]{i-histo_125.eps}
\centering \includegraphics[angle=-90,scale=0.35]{z-histo_99.eps}
\centering \includegraphics[angle=-90,scale=0.35]{W1-histo_200.eps}
\centering \includegraphics[angle=-90,scale=0.35]{W2-histo_200.eps}
\caption{The distribution of each of the magnitudes for various imputation methods.
Note that the un-imputed, maximum, mean, median and most frequent distributions are coincident
apart from the spikes introduced by these methods.}
\label{SDSS_histos}
\end{figure*}

Given that only 72\% of the available training data are complete, we 
explored various methods for replacing (imputing) the missing data:
\begin{enumerate}
\item {\em Univariate (single) imputation: } The most straightforward  method, which replaces the
missing magnitude with a single value. This can be done via:
\begin{enumerate}
\item {\em Simple imputation: } Replacing the missing value with the mean, median or most frequently occuring value. An arbitrary constant may also be 
assigned. 
\item {\em Maximum value imputation: } The  missing value is replaced by the detection limit \citep{cma+21}, which we implemented 
by  assuming that this  was given by the largest value of the magnitude in question.
\item {\em Hot-deck imputation:} The missing magnitudes are replaced by random values. 
We did this by generating a 
random number with a value between the minimum and maximum of that particular magnitude. 
\end{enumerate}
\item {\em Model-based imputation:} Each missing value is modelled
using the other non-missing features of the dataset.  A feature column (magnitude) is the
output, with the remainder of the magnitudes acting as the inputs \citep{lr86,vg11}. We tested:
\begin{enumerate} 
\item {\em Multivariate  (multiple)  imputation:}  A regression is fit for a known output and
used to predict the empty values, which is then iterated for each feature until the maximum specified number of
iterations is reached.
We used the 
{\tt IterativeImputer} function of {\sf sklearn}\footnote{https://scikit-learn.org/stable/}, 
with a maximum number of iterations set to 1000, in order to reach
the early stopping criterion.
\item {\em k-nearest neighbours (kNN) imputation:} For each missing value, the Euclidean distance is found for 
$k$ nearest neighbours for which the feature has a value.
 The neighbour features can either be weighted 
uniformly or by the inverse of the Euclidean distance \citep{tcs+01}.  
This was implemented using {\tt  KNNImputer} function of  {\sf sklearn}, for which we found $k\sim10$
nearest neighbours uniform weighting to perform the best
and so we tested this for imputed models which used $k=3,10$ and 20.
\end{enumerate}
\end{enumerate}
Since the aim is to predict the unavailable source redshifts, we removed the redshifts from the data 
before imputing, to ensure that these did not contribute to the machine learning.
We summarise the results in Table~\ref{SDSS_stats} and 
\begin{table*}
\centering
\caption{The statistics of the un-imputed and imputed SDSS QSO data. 
Where the minimum and maximum
are not shown where these are the same as the un-imputed data.}
\begin{tabular}{@{}l  c c c  c c c c  c c c c @{}}
\hline
\smallskip
         &   Redshift &    $FUV$ &    $NUV $  & $u$ &            $g$ &    $r$  &     $i$ &   $z$ &    $W1$ &    $W2$ &    \\
\hline
\multicolumn{11}{c}{  UN-IMPUTED (missing values included)}\\ 
\hline
$n$  &  100\,337 & 89\,679 &  80\,037  & 97\,414  & 99\,036 & 99\,107 &   99\,081&  99\,079 & 96\,297 & 96\,181 \\
mean    & 1.483 & 20.81  & 20.57 & 19.76 & 19.39  & 19.14  &    18.98 & 18.91 &    15.16 & 14.21 \\
std         & 0.878 &    0.95 &     1.26  & 1.27 &   0.19  & 0.77 &     0.74 & 0.75 & 0.91   & 0.97 \\
min        & 0.0046 &   16.07&  15.25  & 15.38 & 14.47 &   14.38 &   13.61 & 13.74 &    9.91 &  8.90 \\
max     & 6.999 &     24.92 &     25.53  & 31.85 & 29.32 & 22.62 &    27.17 & 24.26 & 27.84 &     27.70 \\
\hline
\multicolumn{11}{c}{{ UNIVARIATE (mean),} $n = 100\,337$}\\
\hline
mean      &  --- &  20.81 &  20.57 & 19.76& 19.39 &  19.14 & 18.98 &  18.91 & 	15.16 &  14.21 \\ 
std         &  --- &     0.90 &      1.13 &       1.25 &       0.89 &     0.76 &   0.74 &   0.75&      	  0.89&     0.95\\  
\hline
\multicolumn{11}{c}{{ UNIVARIATE (median),} $n = 100\,337$}\\ 
\hline
mean      &  --- &  20.81 &    20.55 &  19.75& 	 19.39 &       19.14 &     18.98 &     18.91 & 		15.16&      14.21 \\	
std         &  --- &     0.90 &      1.13&      1.25 &        0.89 &      0.76&        0.74 &        0.75 & 0.89&  0.95\\
\hline
\multicolumn{11}{c}{{ UNIVARIATE (most frequent),} $n = 100\,337$}\\
\hline
mean      &  --- &  20.87 &   20.51 &  19.73 &   19.38 &    19.14 &      18.98 &   18.91 &   	 15.16 &      14.20 \\
std         &  --- &    0.91 &       1.14 &       1.26 &     0.89 &       0.7 7& 0.74 &      0.74 &            0.89&      0.95\\
\hline
\multicolumn{11}{c}{{ UNIVARIATE (maximum),} $n = 100\,337$}\\
\hline
mean     &     ---      &    21.25 &  21.58 &   20.11 & 19.52 &  19.18 & 19.09 &    18.97 & 15.67 &  14.77 \\
std        &       ---      &     1.55 & 2.29 &      2.39 & 1.43 & 0.855&  1.17&     0.953 & 2.65 &       2.85 \\
\hline
\multicolumn{11}{c}{{ UNIVARIATE (maximum truncated),} $n = 100\,337$}\\
\hline
mean     &     ---   & 21.25 & 21.58 &     20.11 & 19.52 &    19.18 & 19.03 &      18.97 & 15.29 &       14.35\\
std        &       ---     &  1.55 & 2.29 &       2.39 & 1.43&     0.86 &      0.83 &     0.95 &   1.09 &   1.16\\
max       &    ---         &     24.92 & 25.53 &  31.85 & 29.32 &     22.62 &      23.0 &     24.26 & 19.0&   19.0\\
\hline
\multicolumn{11}{c}{{ UNIVARIATE (random),} $n = 100\,337$}\\
\hline
mean     &     ---      &20.81 & 20.57 & 19.76 & 19.39 &   19.14&  18.98 &  18.91 & 15.16 &   14.21\\
std        &       ---      & 0.95 & 1.26 & 1.27 & 0.90 &   0.77 & 0.74 &  0.75 & 0.91 &    0.97\\
\hline
\multicolumn{11}{c}{{ MULTIVARIATE,} $n = 100\,337$}\\
\hline
mean   &  --- & 20.83 & 20.80 &   19.82 & 19.40 &   19.14 & 18.99 &   18.91 & 15.18&   14.23\\
std        &  ---  & 0.94 &   1.42 & 1.34 &   0.90 &      0.77 &    0.74&       0.75 &  15.18 &     14.23\\
min       &   ---  &  16.07 & 10.21 &    15.38& 14.47 &  14.38 &  13.60 & 13.74 & 9.91 &       8.90\\
 max     &  ---   & 27.41 & 33.22 &     36.82 & 29.32 &   22.62 &     27.17 &      26.22 &   27.84&      27.70\\
\hline
\multicolumn{11}{c}{{ kNN ($k=3$),} $n = 100\,337$}\\
\hline
mean   &  --- &   20.83& 20.59 &  19.84 &19.40 &  19.14 & 18.99 &  18.91 &  15.17 &    14.23 \\
std        &  ---  & 0.94 & 1.21 &  1.37&    0.90&     0.77 &   0.74 &     0.75 &               0.90&    0.97\\	 
\hline
\multicolumn{11}{c}{{ kNN ($k=10$),} $n = 100\,337$}\\
\hline
mean   &  --- &    20.82 &   20.62 & 19.84 &   19.40   & 19.14 &   18.99 &      18.91&   15.17&  14.23\\
std        &  ---  &  0.93 &     1.19&  1.37 &        0.90&  0.77 &    0.74&    0.75 &   	0.90&      0.97\\
\hline
\multicolumn{11}{c}{{ kNN ($k=20$),} $n = 100\,337$}\\
\hline
mean   &  --- &    20.82& 20.65 &  19.84 & 19.40 &     19.14 &     18.98	&18.91  &   15.17&     14.23 \\	      
std        &  ---  & 0.92 &   1.10 &    1.36& 	0.90 &     0.77 &      0.74& 	 0.75 &     0.90 &       0.97\\   
\hline
\end{tabular}
\label{SDSS_stats}  
\end{table*} 
in Fig.~\ref{SDSS_histos}  we show the distributions of the un-imputed\footnote{We use {\em un-imputed} to refer to the full 100\,337 sample
with the missing values included and {\em non-imputed} for the 72\,276  sources with all nine magnitudes.} and
imputed magnitudes. From the figure, we see that, as expected, the imputed non-SDSS photometry
showed larger deviations from the measured magnitudes, most likely due to the larger 
number of missing values (Table~\ref{missing}). However, this only appears to be particularly
severe for the $NUV$ data, which have the most missing values.
We also note that the maximum values in some bands may be
due to outliers.
\begin{figure}
\centering \includegraphics[angle=-90,scale=0.55]{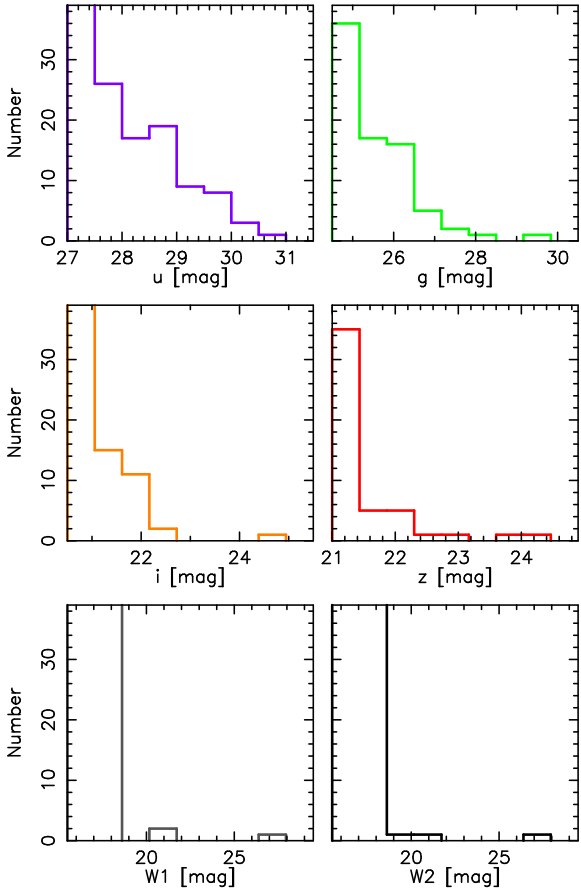}
\caption{The upper extremes of the magnitudes of the un-imputed data with suspected outliers.}
\label{SDSS_outliers}
\end{figure} 
Showing the top ends of the distributions in detail (Fig.~\ref{SDSS_outliers}), we see that this certainly
appears to be the case for the $i$ and WISE bands. We
therefore apply the maximum value imputation, but limiting $i\leq23$ and $W1 \leq 19 \text{ and }W2\leq 19$
({\em maximum truncated} in Table~\ref{SDSS_stats}).

\subsubsection{Deep Learning}
\label{dl}

As described in \citet{cmp21}, the Deep Learning model, built with the 
{\sf TensorFlow}\footnote{https://www.tensorflow.org} platform,
outperformed both the {\em $k$-Nearest Neighbour} (kNN) and {\em Decision Tree Regression} algorithms, with self
validation giving a regression coefficient of the least-squares linear fit between the predicted and measured redshifts of $r=0.94$. 
As with other studies, we used the colours as features (Sect.~\ref{intro}), although we noted that using the raw magnitudes 
gave a similar performance. Since replacing missing magnitudes is more straightforward than colours, we
use the magnitudes as features here.

Upon imputing the data, we trained the model on a random 80\% portion, using the same Deep
Learning model as previously --  two {\em Rectified Linear Unit} (ReLu) function layers and
one {\em hyperbolic tangent} (tanh) layer comprising 200 neurons each. 
We then validated the model on the remaining 20\% of the data, comparing the photometric (predicted)
redshifts with the spectroscopic (measured) redshifts, which had been returned to the imputed dataset.
 In order to account for the small differences
in the results inherent between each trial, we then reshuffled the data and repeated the process 99 times.
We summarise the results in Table~\ref{SDSS_TF},
\begin{table*}
\centering
 \caption{The mean results of the SDSS validation (20\,067 sources) for the  various imputation methods over randomized 100 trials, quoted with $\pm1\sigma$. For the non-imputed data there are 14\,253 validation sources.}
\begin{tabular}{@{}l  c  r c c r c c   @{}}
\hline
\smallskip
Imputation & $r$ & \multicolumn{3}{c}{Un-normalised} & \multicolumn{3}{c}{Normalised}\\
      & & $\mu_{\Delta z}$  & $\sigma_{\Delta z}$ & $\sigma_{\text{MAD}}$ &  $\mu_{\Delta z (\text{norm})}$  & $\sigma_{\Delta z (\text{norm})}$ & $\sigma_{\text{NMAD}}$\\
\hline 
None  & $0.934\pm0.002$ & $0.003\pm0.032$ & $0.234\pm0.003$ & $0.105\pm0.012$ & $-0.008\pm0.015$ & $0.115\pm0.005$ & $0.048\pm0.006$ \\
Univariate    & \multicolumn{7}{c}{} \\
\multicolumn{1}{r}{Mean} & $0.918\pm0.002$  & $0.003\pm0.034$ & $0.351\pm0.004 $ & $0.132\pm0.012$ & $-0.017\pm0.017$ & $0.162\pm0.006$ & $0.055\pm0.005$ \\
\multicolumn{1}{r}{Median} & $0.919\pm0.002$ & $0.002\pm0.032$ & $0.349\pm0.004$ & $0.129\pm0.011$ & $-0.016\pm0.015$ & $0.162\pm0.006$ & $0.054\pm0.005$ \\
\multicolumn{1}{r}{Most} & $0.918\pm0.002$ & $0.002\pm0.032$ & $0.352\pm0.004$ & $0.130\pm0.010$ & $-0.016\pm0.015$ & $0.166\pm0.006$ & $0.055\pm0.004$ \\
\multicolumn{1}{r}{Max} & $0.927\pm0.002$ & $0.004\pm0.035$ & $0.332\pm0.005$ & $0.124\pm0,015$ & $-0.014\pm0.017$ & $0.155\pm0.009$ & $0.051\pm0.007$\\
\multicolumn{1}{r}{Max trun.} & $0.933\pm0.003$ & $0.003\pm0.037$ & $0.315\pm0.007$ & $0.127\pm0.017$ & $-0.014\pm0.016$ & $0.154\pm0.009$ & $0.052\pm0.005$\\
\multicolumn{1}{r}{Random} & $0.909\pm0.002$ & $0.006\pm0.032$ & $0.369\pm0.005$ & $0.134\pm0.011$ & $-0.166\pm0.015$ & $0.175\pm0.007$ & $0.056\pm0.004$\\
Multivariate  & $0.916\pm0.002$  & $0.000\pm0.032$ & $0.355\pm0.005$ & $0.130\pm0.012$ & $-0.018\pm0.015$ & $0.167\pm0.008$ & $0.055\pm0.005$\\
kNN   & \multicolumn{7}{c}{} \\
\multicolumn{1}{c}{$k=3$} & $0.903\pm0.002$ & $-0.004\pm0.030$ & $0.380\pm0.004$ & $0.138\pm0.009$ & $-0.022\pm0.013$ & $0.180\pm0.007$& 
$0.058\pm0.004$\\
\multicolumn{1}{c}{$k=10$} & $0.909\pm0.002$ & $-0.003\pm0.029$ & $0.369\pm0.004$ & $0.132\pm0.011$ & $-0.021\pm0.013$ & $0.175\pm0.006$ & $0.056\pm0.005$ \\
\multicolumn{1}{c}{$k=20$} & $0.912\pm0.002$ & $0.001\pm0.026$& $0.364\pm0.004$ & $0.129\pm0.010$ & $-0.019\pm0.012$ & $0.172\pm0.007$ & $0.054\pm0.004$\\
\hline
\end{tabular}
\label{SDSS_TF}  
\end{table*}
using statistics calculated for the difference between the photometric and spectroscopic redshifts,
$\Delta z \equiv z_{\rm spec} - z_{\rm phot}$, and its normalized counterpart, 
$\Delta z(\text{norm}) \equiv \frac{z_{\rm spec} - z_{\rm phot}}{z_{\rm spec} +1}$.
Specifically, the mean difference between the  photometric and spectroscopic redshifts, 
\[
\mu_{\Delta z} \equiv \frac{1}{N}\sum_{i=1}^N \Delta z, 
\]
the standard deviation, 
\[
\sigma_{\Delta z} \equiv \sqrt{\frac{1}{N}\sum_{i=1}^N \Delta z^2},
\]
and the median absolute deviation (MAD),
\[
\sigma_{\text{MAD}} \equiv 1.48 \times \text{median}\left|z_{\rm spec} - z_{\rm phot}\right|.
\]

We confirm that using the magnitudes directly performs as well as 
 the $FUV - NUV$, $NUV-u$, $u- g$, $g - r$, $r
- i$, $i - z$, $z-W1$ \& $W1-W2$ colours, where the regression coefficient was $r\approx0.94, \sigma_{\Delta  z} \approx 0.235$ and $\sigma_{\text{MAD}}\approx 0.092$ \citep{cmp21}.\footnote{These are given as
approximations since they are from a single run of the model.}  
We also see that the best performing
imputation method is the {\em maximum truncated}, with $\overline{r}= 0.933, \overline{\sigma_{\Delta z}} = 0.315$ and
$\overline{\sigma_{\text{MAD}}} = 0.127$.  This uses a  similar  method as \citet{cma+21}, who assume the detection limit for
each missing value. According to their chosen metric, the {\em normalised median absolute
  deviation} (NMAD),
\[
\sigma_{\text{NMAD}} \equiv 1.48 \times \text{median}\left|\frac{z_{\rm spec} - z_{\rm phot}}{z_{\rm spec} +1}\right|,
\]
our model performs slightly better than theirs (all have $\sigma_{\text{NMAD}} >0.06$, cf. our 0.052). As previously noted
\citep{cmp21}, Deep Learning gives better results across all metrics, including
$\sigma_{\text{NMAD}}$, than standard 
machine learning algorithms (namely, $k$-Nearest Neighbour and Decision Tree Regression).

\subsection{The test data}
\label{test_data}

\subsubsection{The data}
\label{td}

\begin{figure}
\hspace*{-0.2cm}
\centering \includegraphics[angle=0,scale=0.7]{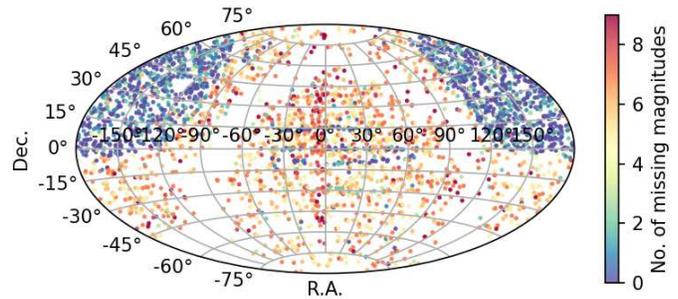}
\caption{The sky distribution of the OCARS sources, colour coded by the number of missing
magnitudes per source.}
\label{sky}
\end{figure}

\begin{figure}
\centering \includegraphics[angle=-90,scale=0.5]{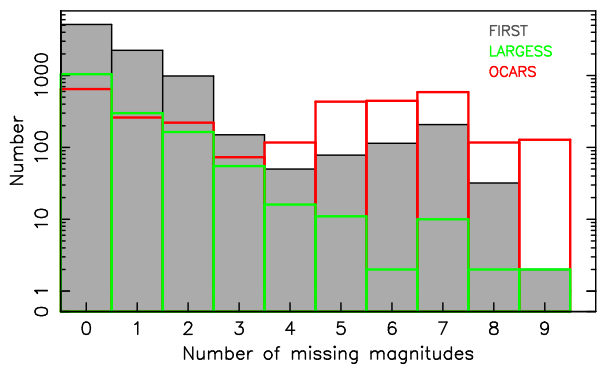}
\caption{Histograms of the total number of missing magnitudes per source for the three test samples.}
\label{hist_missing}
\end{figure}

As stated above, our goal is to develop photometric redshift models for
 radio sources detected with the SKA and its pathfinders. As described in \citet{cm19},
finding large catalogues of
radio sources with spectroscopic redshifts is a challenge and just three sizable databases were found:
\begin{enumerate}

\item  The {\em Faint Images of the Radio Sky at Twenty-Centimeters} (FIRST, \citealt{bwh95,wbhg97}), which has
18\,273 sources with redshifts from the SDSS DR14 QSOs \citep{ppa+18}.  
Of these,  9016 are classed as QSOs in NED. 

\item The {\em Large Area Radio Galaxy Evolution Spectroscopic Survey} (LARGESS). Of the 10\,685 sources with 
  optical redshifts \citep{csc+17}\footnote{Those with redshift reliability flag $q\geq3$, where 
$q=3$  designates ``a reasonably confident redshift'',  and the maximum $q=5$ designates an ``extremely reliable redshift from a good-quality spectrum''.}, 1608
  are classified as quasars in NED. Although termed ``radio-loud'', these have a similar distribution of radio fluxes as the
  SDSS sample \citep{cmp21} and so we refer to these as QSOs.

\item The {\em Optical Characteristics of Astrometric Radio Sources} (OCARS) catalogue of {\em Very Long Baseline
  Interferometry} (VLBI) astrometry sources \citep{mab+09,mal18}. Of the 3663 sources, 2404 are classified as quasars,
but given that these are very strong radio calibration sources we assume that all of the non-galaxies are
active galactic nuclei (AGN), giving a sample size of 3033.
\end{enumerate}
\begin{table}
\centering
\caption{The statistics for the test samples. $N$ is the number of sources for which the magnitude is available,
followed by the percentage of the total. This is followed by the minimum, maximum, mean, median and most frequently measured value of the magnitude.}
\begin{tabular}{@{}l  r r r r r r r @{}}
\hline
\smallskip
Mag.     & $N$ & \% & min & max & mean & median & most\\
\hline
\multicolumn{8}{c}{FIRST, $n=9016$} \\
\hline 
$FUV$ &   6912 & 77 & 16.07 & 25.03 & 20.78 &20.85 & 21.38 \\
$NUV$ &  6699 & 74  & 14.83 & 25.40 & 20.57 & 20.52 & 21.43\\
$u$ &     8457& 94 & 14.22 & 28.45 & 19.76 & 19.65 & 18.88 \\
$g$ &    8672& 95 & 14.49 & 26.34 & 19.30 &19.32 & 18.87 \\
$r$ &     8587 & 95 & 14.38 & 22.18 & 19.02 & 19.07 &  18.87\\
$i$ &    8564 & 95 & 14.45 & 22.06 & 18.85 &    18.92 &  18.85\\
$z$ &    8563 & 95 & 14.35 & 22.55 & 18.75& 18.83 & 18.81 \\
$W1$ &  8563& 95 & 9.84 & 18.34 & 14.96 & 15.05 & 15.84\\
$W2$ & 8552 & 95 & 8.90&  17.97  & 14.05 & 14.14 & 14.81\\
\hline
\multicolumn{8}{c}{LARGESS, $n=1608$} \\
\hline
$FUV$ &   1265  & 79 & 16.02 &  24.40 & 21.23 & 21.23 & 20.27 \\
$NUV$ & 1373 & 85 & 15.79 &  24.92 & 20.96 & 21.02 & 19.73 \\
$u$ &   1549 &     96 &   15.65 & 30.36 &  20.13 &20.10 & 20.97\\
$g$ &   1583 &  98&  14.27  & 24.16 & 19.67 & 19.72 & 20.38\\
$r$ &   1582  & 98& 14.09 & 22.21  & 19.34 & 19.43 & 18.74 \\
$i$ &   1579  &  98& 13.61 & 21.49 & 19.14 &  19.26 & 18.85\\
$z$ & 1579  &  98& 13.61 & 23.30 & 19.01 & 19.13 & 18.81\\
$W1$ & 1468 &  91 & 10.12&  17.77 & 15.02 & 15.16 & 14.61 \\
$W2$ &  1466 &   91& 8.89 &  17.04 & 14.17 & 14.32 & 15.15\\
\hline
 \multicolumn{8}{c}{OCARS, $n=3033$} \\
\hline  
$FUV$ &   1434&  47 & 13.51& 24.90 & 20.22&  20.31 & 20.82\\ 
$NUV$ & 2035 &   67& 13.56 &24.66 & 20.01&  20.01 & 21.15\\  
$u$ & 1170 &39  & 12.78 & 27.47 & 19.30&  19.18 & 18.26\\ 
$g$ & 1255  &  41& 12.45 &  26.34 & 18.86&  18.78 & 18.74 \\ 
$r$ & 1374 & 45 & 12.77 &  26.28 & 18.57&  18.55 & 18.75\\ 
$i$ & 1264 & 42 & 11.67 &  24.09 & 18.45& 18.48 & 18.70 \\ 
$z$ & 1205 & 40  & 12.69 &  23.13 & 18.30 & 18.34& 18.61 \\
$W1$ & 2562& 84 & 7.26 & 17.56&  14.20 & 14.33 & 13.77\\ 
$W2$ & 2562 & 84 & 6.26 & 17.23 & 13.30 &  13.39 & 13.54 \\
\hline
\end{tabular}
\label{test_stats}  
\end{table} 
Again, the requirement of measured magnitudes in all nine bands
has the effect of cutting the samples, leaving 
65\% of the LARGESS sample, 57\% of the FIRST sample and just 21\% of the OCARS sample.
The large fraction of missing data in the latter is due a mismatch between this and the SDSS's
sky coverage, which is restricted to the northern sky (Fig.\ref{sky}).
In Fig.~\ref{hist_missing}, we show 
distribution of the number of missing magnitudes per source, from which we see that OCARS is
most  likely to have 5--7
magnitudes missing, in addition to a large proportion of sources having no magnitude measurements at all.

\subsubsection{Data imputation and machine learning}

We impute the missing values using the same methods as for the SDSS data (Sect.~\ref{sec:methods}). The only method we
exclude is the maximum truncated, since the maximum values for the test data do not indicate the presence of extreme
outliers (Table~\ref{test_stats}).  Following the imputation and upon returning the spectroscopic
redshifts to the data, we train the model using the non-imputed SDSS photometry
(72\,276 QSOs), which we then validate on each test sample, again comparing the predicted with the measured
redshifts (Table~\ref{all_imp}).
\begin{table*}
\centering
 \caption{The mean results for the various imputation methods over 10 trials, quoted with $\pm1\sigma$.}
\begin{tabular}{@{}l c c  r c c r c c   @{}}
\hline
\smallskip
Imputation & $n$ & $r$ & \multicolumn{3}{c}{Un-normalised} & \multicolumn{3}{c}{Normalised}\\
    &     & & $\mu_{\Delta z}$  & $\sigma_{\Delta z}$ & $\sigma_{\text{MAD}}$ &  $\mu_{\Delta z (\text{norm})}$  & $\sigma_{\Delta z (\text{norm})}$ & $\sigma_{\text{NMAD}}$\\
\hline
\multicolumn{9}{c}{FIRST} \\
\hline
None  & 5147 & $0.949\pm0.004$ & $0.031\pm0.019$& $0.206\pm0.006$ & $0.116\pm0.005$ & $0.008\pm0.009$ & $0.094\pm0.002$ & $0.052\pm0.002$ \\
Univariate    & \multicolumn{8}{c}{} \\
\multicolumn{1}{r}{Mean} &  9016 & $0.752\pm0.014$ & $0.109\pm0.052$ & $0.554\pm0.005$ & $0.176\pm0.014$ & $0.014\pm0.024$ & $0.202\pm0.009$ & $0.075\pm0.007$ \\
\multicolumn{1}{r}{Median} &   9016 & $0.740\pm0.033$ & $0.121\pm0.039$ & $0.564\pm0.032$ & $0.169\pm0.012$ & $0.018\pm0.015$ & $0.206\pm0.005$ & $0.072\pm0.005$ \\
\multicolumn{1}{r}{Most}  &  9016 & $0.683\pm0.048$ & $0.180\pm0.054$ & $0.611\pm0.038$ & $0.184\pm0.010$ & $0.035\pm0.021$ & $0.208\pm0.011$ & $0.081\pm0.005$ \\
\multicolumn{1}{r}{Max} &   9016 & $0.771\pm0.012$ & $-0.141\pm0.026$ & $0.597\pm0.028$ & $0.172\pm0.009$ & $-0.078\pm0.009$ & $0.299\pm0.018$ & $0.070\pm0.004$ \\
\multicolumn{1}{r}{Random} & 9016 & $0.680\pm0.030$ & $0.026\pm0.072$ & $0.648\pm0.025$ & $0.191\pm0.009$ & $-0.024\pm0.027$ & $0.276\pm0.108$ & $0.081\pm0.005$ \\
 Multivariate & 9016 & $0.904\pm0.005$ & $0.020\pm0.025$ & $0.359\pm0.010$ & $0.146\pm0.009$ & $-0.004\pm0.011$ & $0.155\pm0.006$ & $0.060\pm0.001$ \\
kNN   & \multicolumn{8}{c}{} \\
 \multicolumn{1}{c}{$k=3$} & 9016 & $0.839\pm0.017$ & $0.069\pm0.051$ & $0.046\pm0.021$ & $0.159\pm0.014$ & $0.009\pm0.021$ & $0.170\pm0.003$ & $0.066\pm0.006$ \\
\multicolumn{1}{c}{$k=10$} & 9016 & $0.872\pm0.012$ & $0.054\pm0.027$ & $0.409\pm0.019$ & $0.149\pm0.007$ & $0.006\pm0.004$ & $0.160\pm0.005$ & $0.063\pm0.001$ \\
\multicolumn{1}{c}{$k=20$} & 9016 & $0.890\pm0.007$ & $0.022\pm0.028$ & $0.382\pm0.011$ & $0.144\pm0.006$ & $-0.006\pm0.013$ & $0.155\pm0.005$ & $0.060\pm0.002$ \\
\hline
\multicolumn{9}{c}{LARGESS} \\
\hline
 None  & 1046 & $0.913\pm0.005$ & $0.042\pm0.022$ & $0.292\pm0.008$ & $0.128\pm0.005$ & $0.008\pm0.012$ & $0.116\pm0.009$ & $0.061\pm0.002$ \\
Univariate    & \multicolumn{8}{c}{} \\
\multicolumn{1}{r}{Mean} &  1608 & $0.720\pm0.020$ & $0.080\pm0.053$ & $0.610\pm0.022$ & $0.172\pm0.011$ & $-0.005\pm0.023$ & $0.271\pm0.031$ & $0.077\pm0.006$ \\
\multicolumn{1}{r}{Median} &   1608 & $0.723\pm0.012$ & $0.075\pm0.033$ & $0.605\pm0.012$ & $0.175\pm0.020$ & $-0.008\pm0.018$ & $0.271\pm0.021$ & $0.079\pm0.010$ \\
\multicolumn{1}{r}{Most}  &  1608 & $0.641\pm0.024$ & $0.105\pm0.049$ & $0.683\pm0.025$ & $0.196\pm0.020$ & $-0.005\pm0.022$ & $0.305\pm0.025$ & $0.088\pm0.009$ \\
\multicolumn{1}{r}{Max} & 1608 & $0.809\pm0.013$ & $-0.036\pm0.026$ & $0.531\pm0.027$ & $0.171\pm0.009$ & $-0.041\pm0.013$ & $0.297\pm0.021$ & $0.074\pm0.003$ \\
\multicolumn{1}{r}{Random} & 1608 & $0.621\pm0.025$ & $0.024\pm0.046$ & $0.742\pm0.040$ & $0.190\pm0.008$ & $-0.034\pm0.019$ & $0.361\pm0.029$& $0.085\pm0.005$ \\
 Multivariate & 1608 & $0.829\pm0.012$ & $0.028\pm0.025$ & $0.488\pm0.017$ & $0.157\pm0.006$ & $-0.013\pm0.011$ & $0.224\pm0.021$ & $0.068\pm0.002$ \\
kNN   & \multicolumn{8}{c}{} \\
\multicolumn{1}{c}{$k=3$} &  1608 & $0.814\pm0.010$ & $0.023\pm0.024$ & $0.511\pm0.018$ & $0.162\pm0.007$ & $-0.017\pm0.012$ & $0.249\pm0.019$ & $0.071\pm0.003$ \\
\multicolumn{1}{c}{$k=10$} & 1608 & $0.816\pm0.016$ & $0.031\pm0.028$ & $0.506\pm0.026$ & $0.013\pm0.007$ & $-0.013\pm0.015$ & $0.242\pm0.027$ & $0.070\pm0.003$ \\
\multicolumn{1}{c}{$k=20$} &1608 & $0.811\pm0.010$ & $0.044\pm0.048$ & $0.513\pm0.022$ & $0.161\pm0.007$ & $-0.009\pm0.021$ & $0.238\pm0.034$ & $0.071\pm0.003$ \\
\hline
\multicolumn{9}{c}{OCARS} \\
\hline
None  &  649 & $0.867\pm0.010$ & $0.048\pm0.043$ & $0.339\pm0.003$ & $0.144\pm0.013$ & $0.008\pm0.025$ & $0.159\pm0.014$ & $0.066\pm0.007$ \\
 Univariate    & \multicolumn{8}{c}{} \\
\multicolumn{1}{r}{Mean} &  3033 & $0.482\pm0.020$ & $0.181\pm0.015$ & $0.823\pm0.012$ & $0.476\pm0.017$ & $0.015\pm0.015$ & $0.337\pm0.009$ & $0.218\pm0.009$ \\
\multicolumn{1}{r}{Median} &   3033 & $0.483\pm0.011$ & $0.114\pm0.055$ & $0.821\pm0.013$ & $0.478\pm0.016$ & $-0.017\pm0.026$ & $0.350\pm0.011$ & $0.217\pm0.008$ \\
\multicolumn{1}{r}{Most}  &  3033 & $0.421\pm0.018$ & $0.190\pm0.047$ & $0.807\pm0.011$ & $0.518\pm0.022$ & $-0.002\pm0.008$ & $0.224\pm0..08$ & $0.238\pm0.006$ \\
\multicolumn{1}{r}{Max} & 3033 & $0.346\pm0.048$ & $0.281\pm0.165$ & $0.943\pm0.049$ & $0.638\pm0.060$ & $0.041\pm0.073$ & $0.411\pm0.032$ & $0.313\pm0.039$ \\
\multicolumn{1}{r}{Random} & 3033 & $0.282\pm0.014$ & $0.070\pm0.100$ & $1.043\pm0.040$ & $0.720\pm0.039$ & $-0.055\pm0.046$ & $0.478\pm0.024$ & $0.321\pm0.018$ \\
Multivariate &   3033 & $0.597\pm0.011$ & $0.240\pm0.050$ & $0.722\pm0.009$ & $0.379\pm0.013$ & $0.050\pm0.024$ & $0.284\pm0.008$ & $0.183\pm0.007$ \\
kNN   & \multicolumn{8}{c}{} \\
\multicolumn{1}{c}{$k=3$} & 3033 & $0.506\pm0.008$ & $0.238\pm0.040$ & $0.813\pm0.011$ & $0.466\pm0.011$ & $0.043\pm0.018$ & $0.326\pm0.013$ & $0.221\pm 0.004$ \\
\multicolumn{1}{c}{$k=10$} & 3033 & $0.513\pm0.017$ & $0.314\pm0.032$ & $0.785\pm0.010$ & $0.438\pm0.018$ & $0.078\pm0.015$ & $0.295\pm0.002$ & $0.215\pm0.009$\\ 
\multicolumn{1}{c}{$k=20$} &3033 & $0.540\pm0.026$ & $0.310\pm0.028$ & $0.767\pm0.015$ & $0.444\pm0.023$ & $0.077\pm0.017$ & $0.290\pm0.008$ & $0.212\pm0.011$\\
\hline
\end{tabular}
\label{all_imp}  
\end{table*} 
We see that replacing the missing magnitudes via machine learning methods (multivariate and kNN imputation)
significantly outperform the simple  (univariate and random) methods, although replacement with the maximum value
remains the best of these. The effectiveness of applying the maximum value to the SDSS data likely stems
from many of the missing values actually being due to sensitivity limits, whereas for data drawn independently
of the SDSS (i.e. OCARS), a larger fraction of the missing data are expected to be due to non-measurements. 

In order to place the OCARS data on a more equal par with the other test samples, we now consider 
only the OCARS sources with at  least one SDSS magnitude, thus being more likely to 
having overlapping coordinates (Table~\ref{stats+SDSS}),
\begin{table}
\centering
\caption{The statistics for the OCARS quasars with at least one SDSS magnitude (1420 sources), cf. Table~\ref{test_stats}.}
\begin{tabular}{@{}l  r r r r r r r @{}}
\hline
\smallskip
  Mag.     & $n$ & \% & min & max & mean & median & most\\  
\hline  
$FUV$ &   946 &  67 & 13.51& 24.90 & 20.09&  20.15 &   20.39 \\ 
$NUV$ & 1059&  75 &   13.57&  24.66 &  19.83&  19.71 & 20.50\\  
$u$ & 1170 &   82 & 12.78 & 27.47 & 19.30&  19.18 & 18.26\\ 
$g$ & 1255  &  88& 12.45 &  26.34 & 18.86&  18.78 & 18.74 \\ 
$r$ & 1374 & 97& 12.77 &  26.28 & 18.57&  18.55 & 18.75\\ 
$i$ & 1264 &89  & 11.67 &  24.09 & 18.45& 18.48 & 18.70 \\ 
$z$ & 1205 & 85  & 12.69 &  23.13 & 18.30&  18.34 & 18.61 \\
$W1$ & 1269 &  89& 8.31 &17.56 & 14.19&  14.33 & 14.65 \\ 
$W2$ & 1268 & 89 & 7.05 & 16.55 & 13.25&  13.34 & 14.78 \\
\hline
\end{tabular}
\label{stats+SDSS}  
\end{table} 
Applying the Deep Learning to this data (Table~\ref{OCARS_imp}),
\begin{table*}
\centering
 \caption{As Table~\ref{all_imp}, but  for the OCARS sources with at least one SDSS magnitude measurement.}
\begin{tabular}{@{}l c c  r c c r c c   @{}}
\hline
\smallskip
Imputation & $n$ & $r$ & \multicolumn{3}{c}{Un-normalised} & \multicolumn{3}{c}{Normalised}\\
    &     & & $\mu_{\Delta z}$  & $\sigma_{\Delta z}$ & $\sigma_{\text{MAD}}$ &  $\mu_{\Delta z (\text{norm})}$  & $\sigma_{\Delta z (\text{norm})}$ & $\sigma_{\text{NMAD}}$\\
\hline
\multicolumn{9}{c}{OCARS ($\geq0$ SDSS match)} \\
\hline
Univariate    & \multicolumn{8}{c}{} \\
\multicolumn{1}{r}{Mean} &  1420 & $0.546\pm0.023$ & $0.232\pm0.060$ & $0.804\pm0.014$ & $0.263\pm0.023$ & $0.039\pm0.028$ & $0.310\pm0.014$ & $0.125\pm0.010$ \\
\multicolumn{1}{r}{Median} &   1420 & $0.544\pm0.022$ & $0.218\pm0.056$ & $0.801\pm0.026$ & $0.267\pm0.011$ & $0.032\pm0.026$ & $0.314\pm0.023$ & $0.128\pm0.006$ \\
\multicolumn{1}{r}{Most}  &  1420 & $0.529\pm0.026$ &  $0.191\pm0.075$ & $0.816\pm0.025$ & $0.320\pm0.015$ & $0.017\pm0.033$ & $0.317\pm0.007$ & $0.153\pm0.008$ \\
\multicolumn{1}{r}{Max} & 1420 & $0.574\pm0.030$ & $-0.032\pm0.058$ & $0.826\pm0.035$ & $0.292\pm0.012$ & $-0.074\pm0.027$ & $0.397\pm0.023$ & $0.127\pm0.006$ \\
\multicolumn{1}{r}{Random } & 1420 & $0.458\pm0.028$ & $0.032\pm0.073$ & $0.923\pm0.056$ & $0.346\pm0.065$ & $-0.060\pm0.033$ & $0.428\pm0.044$ & $0.157\pm0.010$ \\
Multivariate & 1420 & $0.737\pm0.004$ & $0.156\pm0.033$ & $0.635\pm0.011$ & $0.213\pm0.011$ & $0.031\pm0.016$ & $0.248\pm0.014$ & $0.095\pm0.006$ \\ 
kNN   & \multicolumn{8}{c}{} \\
\multicolumn{1}{c}{$k=3$} & 1420 & $0.699\pm0.017$ & $0.174\pm0.059$ & $0.674\pm0.015$ & $0.232\pm0.017$ & $0.034\pm0.029$ & $0.256\pm0.014$ & $0.150\pm0.008$ \\
\multicolumn{1}{c}{$k=10$} & 1420 & $0.715\pm0.014$ & $0.189\pm0.040$ & $0.652\pm0.013$ & $0.214\pm0.009$ & $0.040\pm0.019$ & $0.240\pm0.010$ & $0.099\pm0.004$ \\
\multicolumn{1}{c}{$k=20$} & 1420 & $0.686\pm0.030$ & $0.199\pm0.061$ & $0.679\pm0.030$ & $0.227\pm0.014$ & $0.041\pm0.010$ & $0.248\pm0.013$ & $0.105\pm0.008$ \\
\hline
\end{tabular}
\label{OCARS_imp}  
\end{table*} 
we see again  that the multivariate method is the best, although in general the results are inferior to those of the 
FIRST and LARGESS samples. However, even with the requirement of at least one SDSS magnitude, only 46\% of the OCARS
sources have the full magnitude complement.

\section{Discussion}

\subsection{Results}
\subsubsection{General}

Training and validating on the 100\,337  strong SDSS sample, for which 28\% of sources have at least one missing magnitude,
we find all of the imputation methods to be effective, with the randomly selected 20\,067 validation sources
giving  regression coefficients of $r\gapp0.9$ between the predicted and measured redshifts. These 
compare favourably with $r=0.91 - 0.94$ obtained when using the source colours are features.
The best performing method was replacement by the maximum for the band in question ($r=0.93$),
similar to the imputation method of \citet{cma+21}, who assume that each  missing value is at
the detection limit.

Training on the 72\,276 QSOs which have all nine magnitudes and testing on the three other, radio-selected, catalogues, we
find multivariate imputation to perform the best, although we only achieve $\overline{r} =0.60$ 
for the OCARS sample. However, only 47\% of the sources have an SDSS magnitude, due to the full
sky distribution of OCARS, and considering only these increases the regression coefficient to $\overline{r} =0.74$. This is still low,
which we believe is due to only 46\% of this 47\% having all of the nine magnitudes.

\subsubsection{OCARS}

As discussed above, the machine learning performs poorest when applied to the OCARS data. This may not be 
surprising since, unlike the LARGESS and FIRST data  (\citealt{csc+17,ppa+18}), OCARS is compiled
independently of the SDSS.
Also, the mean radio flux density for the OCARS sources is an order of magnitude higher than the others.
These, unlike OCARS, appear to be  truncated at the lower flux densities  ($S_{\rm radio} \lapp1$~mJy), suggesting
 a flux limitation \citep{cmp21}. 
Hence, 
both the LARGESS and SDSS data may be more representative of future continuum radio surveys, which are expected to be
sensitive to flux densities of $S_{\rm radio} \sim0.1$~mJy (e.g. \citealt{nha+11}). We should, however, not yet rule
out using the SDSS to train the NIR-optical-UV photometry of radio-loud sources.

\begin{figure*}
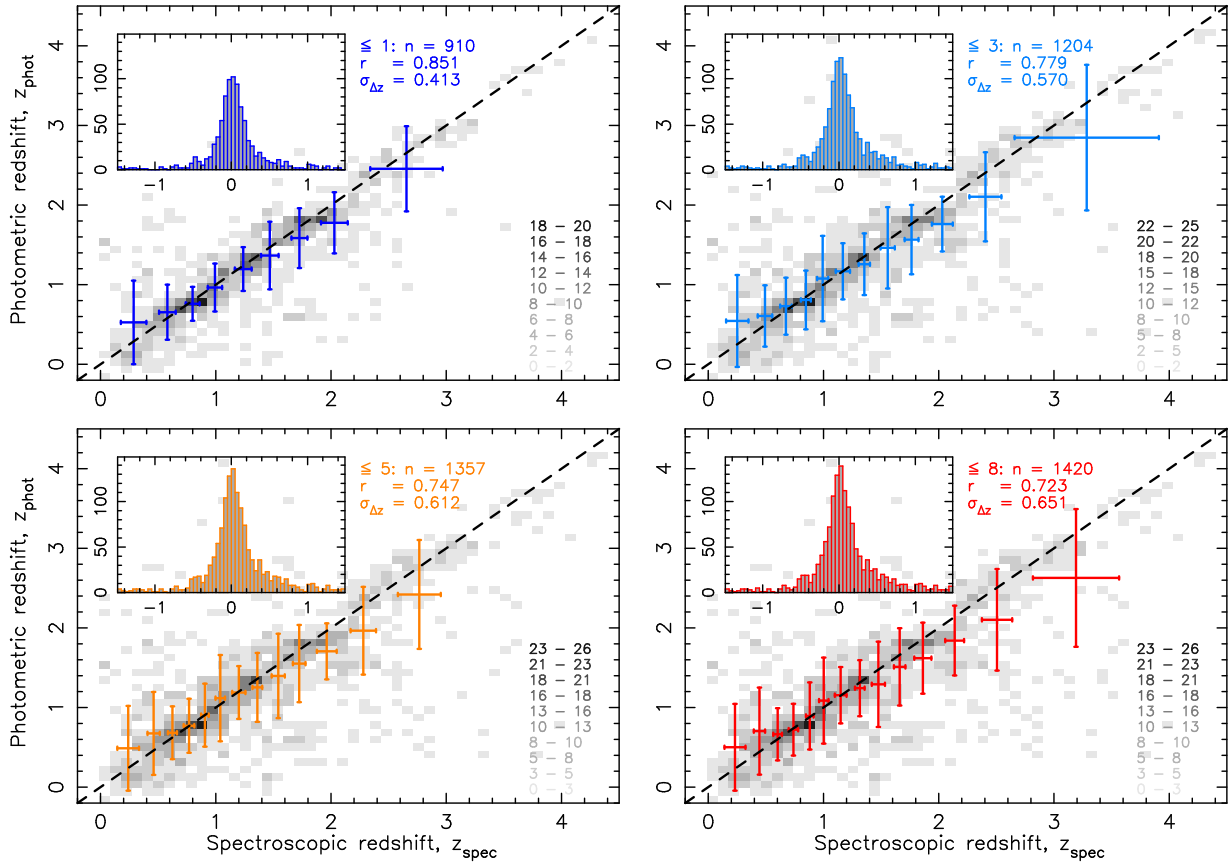
 
\centering \includegraphics[angle=-90,scale=0.5]{pred_spec-grey_histo-equal_1.eps}
\centering \includegraphics[angle=-90,scale=0.5]{pred_spec-grey_histo-equal_3.eps}
\centering \includegraphics[angle=-90,scale=0.5]{pred_spec-grey_histo-equal_5.eps}
\centering \includegraphics[angle=-90,scale=0.5]{pred_spec-grey_histo-equal_8.eps}
\caption{The photometric versus the spectroscopic redshift for different numbers of imputed magnitudes (up to 1,3,5 \& 8) per source
for the OCARS sources with at least one SDSS match. The grey-scale shows the source distribution with the key on the bottom right showing the number within each pixel.
The binning is for an equal number of sources in each bin (100) with the error bars  showing $\pm1\sigma$. 
The broken line shows $z_{\text{phot}} = z_{\text{spec}}$ and the 
inset the distribution of $\Delta z$. Note that this is from a single Deep Learning trial and
so the values should be considered approximate only, e.g. $r = 0.723$ 
for full imputation, cf $\overline{r} = 0.737$ (Table~\ref{OCARS_imp}).}
\label{4z}
\end{figure*}

\begin{figure}
\centering \includegraphics[angle=0,scale=0.32]{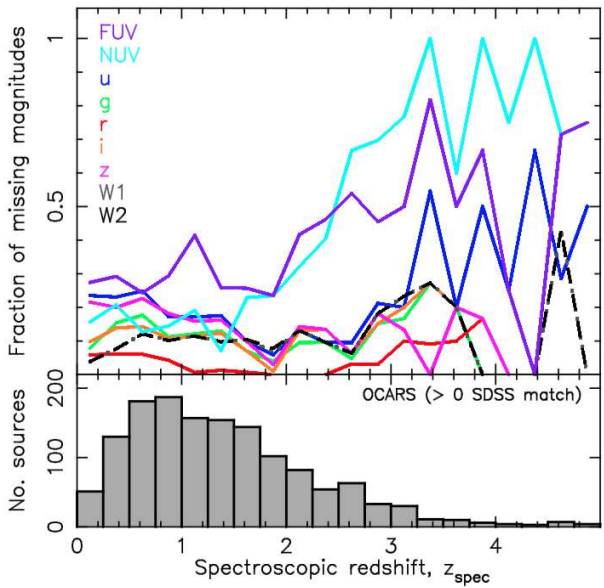}
\caption{The fraction of missing magnitudes in each band (top) and the number of sources (bottom) 
for the OCARS sources with at least one SDSS match.} 
\label{z_obs}
\end{figure}

In Fig.~\ref{4z}, we show how the imputation affects the Deep Learning results for the OCARS quasars as 
more missing magnitudes are replaced. For all cases, we see that the redshift accuracy is relatively poor at 
$z\lapp0.5$, which was also noted for the non-imputed data.
 We \citep{cmp21} suggested that this was due to limitations in the 
FUV magnitudes, which are required to accurately trace  the Lyman break at $z\sim0$ and
in Fig.~\ref{z_obs} we see that the FUV has the largest fraction of missing values at low redshift.
We also see that the fraction of missing values increases across all bands with increasing redshift,
thus requiring a larger degree of imputation. This is evident in Fig.~\ref{4z}
with higher redshifts being attainable with the imputation of more missing values, although at a cost in accuracy.

Imputing all of the missing magnitudes we would retain the whole training sample. Knowing that
the higher redshifts are less accurate, we 
could, in principle, assign a confidence in the prediction dependent upon the 
redshift. But since in practice the redshifts will be unknown a priori, we would require a means of
approximately estimating these. Both \citet{cm19} and \citet{lzc+21} use a two step approach, first
splitting the sources into low and high redshift samples. Specifically, 
\citeauthor{cm19} used the correlation between the $W2$ magnitude and redshift \citep{gas+18}, 
\begin{figure}
\centering \includegraphics[angle=-90,scale=0.5]{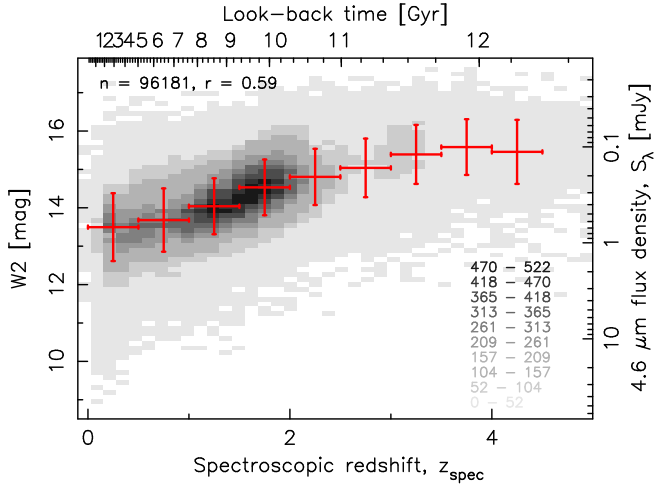}
\caption{The $W2$ magnitude versus redshift for the SDSS sample. 
The binning of the 96\,181 sources is for equal $z_{\text{spec}}$ spacing, with the 
error bars on the abscissa showing the range and the error bars on the ordinate $\pm1\sigma$.}
\label{W2}
\end{figure} 
to obtain an approximate redshift, before proceeding with a more detailed determination.
However, as seen from Fig.~\ref{W2}, this has a large uncertainty and any $W2$--redshift
relation  will already be incorporated 
into the Deep Learning.  

\subsection{Imputation limits}

\begin{figure}
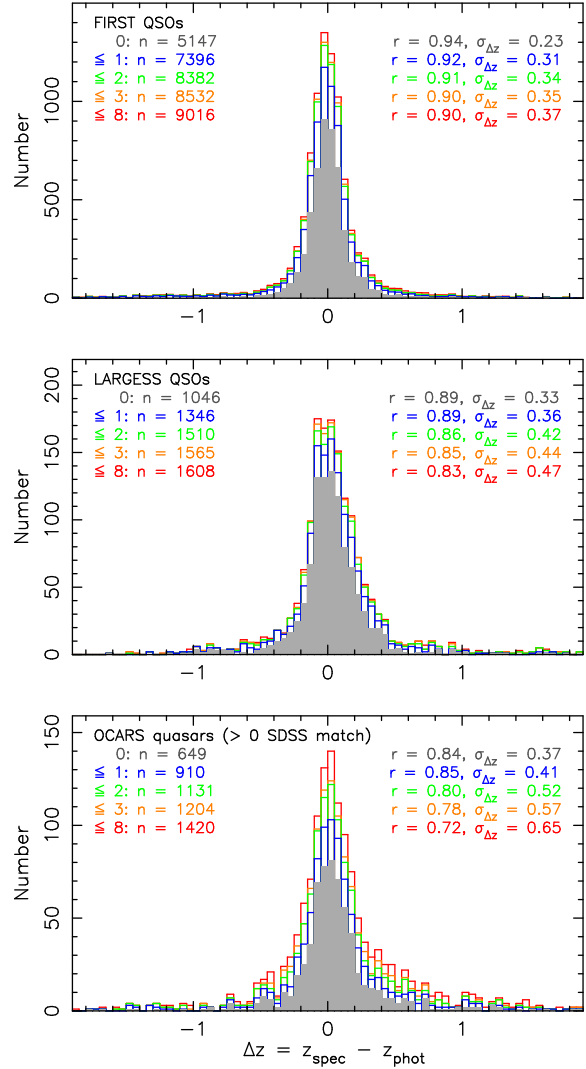

\centering \includegraphics[angle=-90,scale=0.46]{FIRST-hist-bin=76.eps}
\centering \includegraphics[angle=-90,scale=0.46]{LARGESS-hist-bin=76.eps}
\centering \includegraphics[angle=-90,scale=0.46]{OCARS+SDSS-hist-bin=76.eps}
\caption{The spread in $\Delta z$ as more missing magnitudes are imputed for each source using the multivariate method.
From a single  Deep Learning run so the standard deviations are approximate.}
\label{dz_histos}
\end{figure}

In Fig.~\ref{dz_histos}, we show the degradation in the performance as more missing values are replaced for
the three test datasets.
While these are examples of the overall performance, in order to
obtain real insight into how far it is sensible to impute, we require the mean Deep Learning metrics for each number of
missing magnitudes per source: 
\begin{figure*}
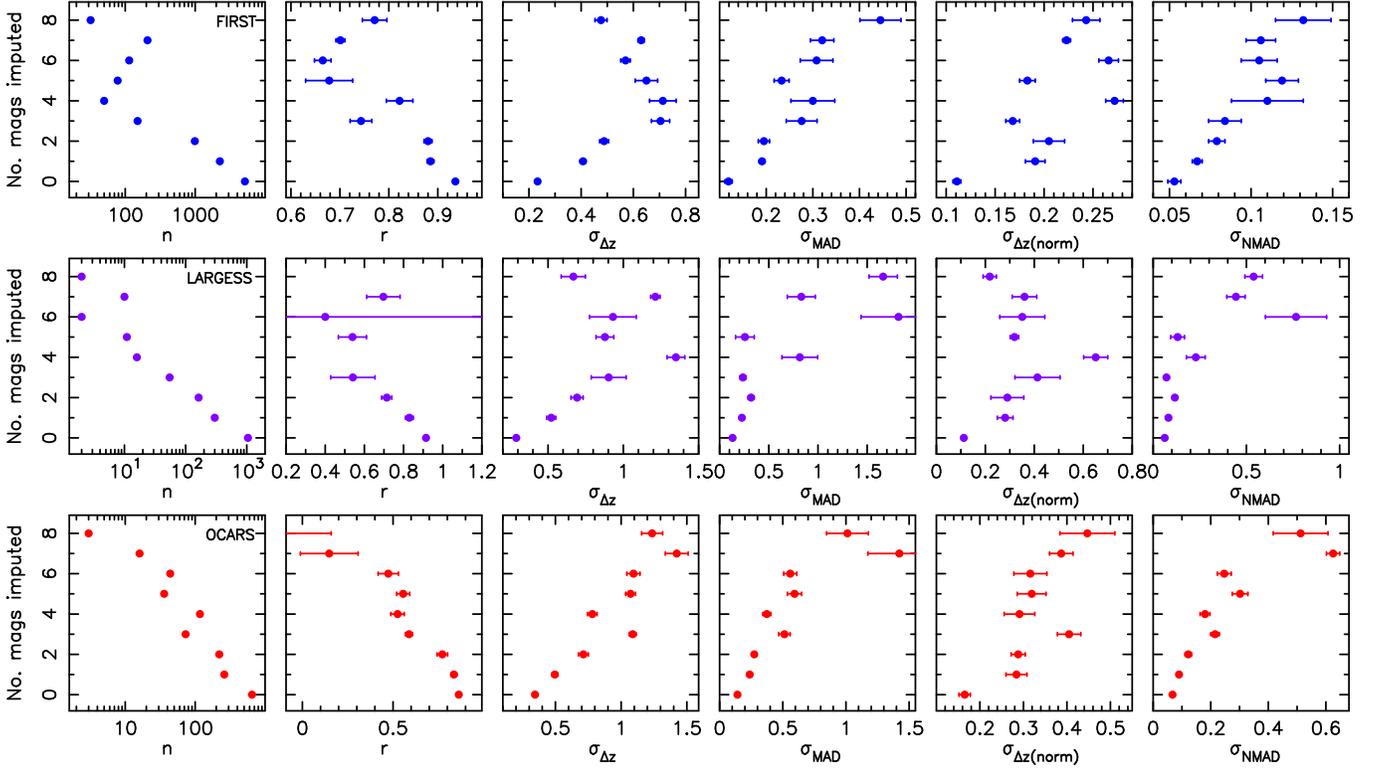

\centering \includegraphics[angle=-90,scale=0.61]{6-FIRST.eps}
\centering \includegraphics[angle=-90,scale=0.61]{6-LARGESS.eps}
\centering \includegraphics[angle=-90,scale=0.61]{6-OCARS.eps}
\caption{The performance of the Deep Learning for different numbers of imputed magnitudes per source
for the FIRST (top) \&  LARGESS (middle) QSOs and OCARS quasars with at least one SDSS magnitude (bottom). 
The error bars show the $\pm1\sigma$ from the mean of 10 trials.}
\label{6-FIRST}
\end{figure*} 
For the FIRST sample (Fig.~\ref{6-FIRST}, top), we see that most sources have less than three missing magnitudes and if 
we limit the imputation to this number,  the predicted redshifts retain a relatively high quality. 

For the LARGESS sample (Fig.~\ref{6-FIRST}, middle)
we see a similar situation, although the decrease in the number of sources as the number of missing values increases is
more gradual. Nevertheless, since we start with a smaller sample, there are only 55 sources with three missing
magnitudes, falling to 16 with four.
Again, the best results cluster at less than three imputed magnitudes per source, with the error bars indicating considerably
less scatter between the Deep Learning runs. Note that there are only two sources with six imputed
magnitudes, resulting in the large errors, e.g. $\overline{r} = 0.4\pm0.9$.

For the OCARS sample (Fig.~\ref{6-FIRST}, bottom), again we see a large drop in performance
when more than two magnitudes are imputed. We note that in
all three cases, the normalised median absolute  deviation (the metric chosen by \citealt{cma+21}) 
appears to the clearest tracer of the degradation of the performance 
as more missing magnitudes are imputed. As discussed in Sect.~\ref{td}, and illustrated in 
Fig.~\ref{sky}, however, around half of the OCARS sources have no SDSS magnitudes due
to their location in the sky. Since the SKA and its pathfinders
are restricted to surveying the southern sky, we can expect very little overlap with the SDSS, thus
not having all of the magnitudes required to predict redshifts. However, {\em SkyMapper} 
\citep{wol+18}\footnote{https://skymapper.anu.edu.au}, which is currently surveying
the southern sky, uses similar bands to the SDSS ($u, v, g, r, i, z$) and so we expect a model
trained upon SDSS data to still be applicable. Of course, once 
the SkyMapper catalogue becomes sufficiently large, the photometry from these sources can also be used to build a model.

In order to demonstrate the performance of the test data with up to two missing magnitudes,
from the original un-imputed data we retain only those with 
less than three missing values before performing multivariate imputation. 
These are then trained using the same  model, created by the 
non-imputed SDSS data, on all three sets. We show the results in 
\begin{figure}
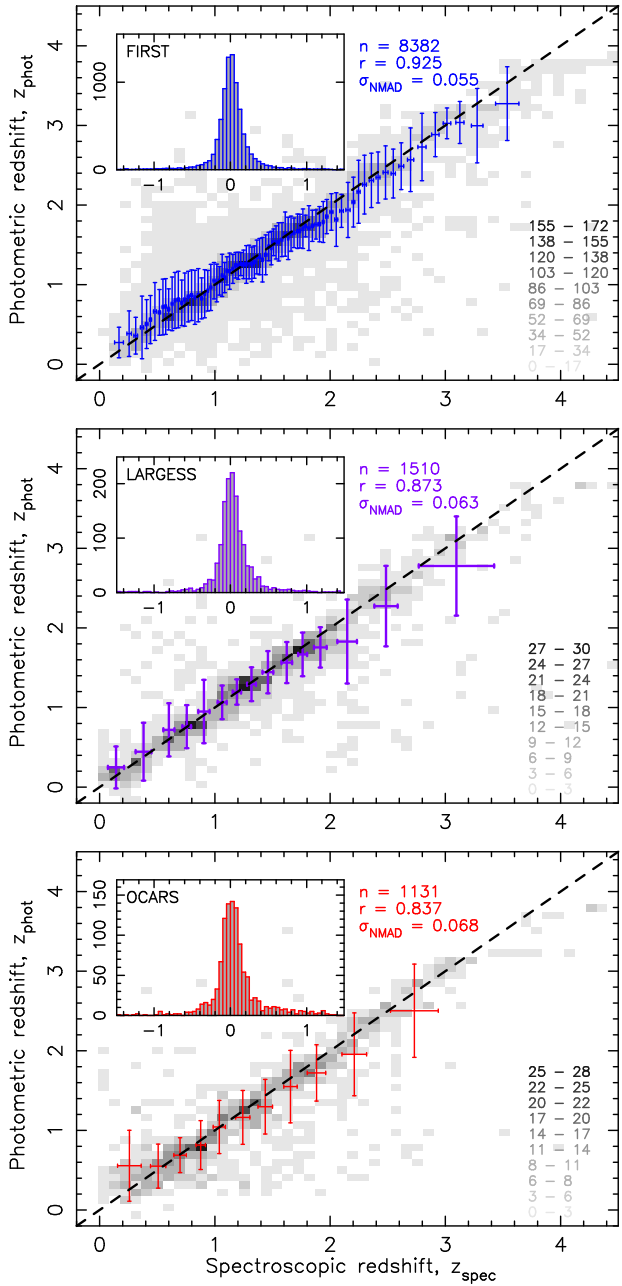
 
\centering \includegraphics[angle=-90,scale=0.5]{pred_spec-FIRST_2.eps}
\centering \includegraphics[angle=-90,scale=0.5]{pred_spec-LARGESS_2.eps}
\centering \includegraphics[angle=-90,scale=0.5]{pred_spec-OCARS_2.eps}
\caption{The results a single Deep Learning run on the test data sets missing up to two magnitudes per
  source.  The binning is for an equal number of sources in each bin (100) with the error bars showing $\pm1\sigma$.  The
  broken line shows $z_{\text{phot}} = z_{\text{spec}}$ and the inset the distribution of $\Delta z$.}
\label{2miss}
\end{figure} 
Fig.~\ref{2miss}, from which we see:
\begin{enumerate}
\item FIRST: $r\approx0.93$ and $\sigma_{\text{NMAD}} \approx0.055$, while retaining 
93\% of the data, compared to $\overline{r} = 0.95$ and $\overline{\sigma_{\text{NMAD}}} = 0.052$ for the 
57\% remaining when the sources with any missing values are excluded.
\item LARGESS:  $r\approx0.87$ and $\sigma_{\text{NMAD}} \approx0.063$ for  94\% of the sample, compared 
to  $\overline{r} = 0.91$ and $\overline{\sigma_{\text{NMAD}}} = 0.061$  for 65\%.
\item OCARS, with at least one SDSS magnitude,  $r\approx0.84$ and  $\sigma_{\text{NMAD}} \approx0.068$ for  
80\%, compared to $\overline{r} = 0.87$ and $\overline{\sigma_{\text{NMAD}}} = 0.066$  for 46\% .
\end{enumerate}
The NMAD values are within the range of  those in the literature, $\sigma_{\text{NMAD}} =0.029$ \citep{ldlr11}, 
0.012 -- 0.058 \citep{bcd+13},  0.060 -- 0.065 \citep{asu+17}, 0.014 -- 0.401 \citep{dbw+18},
$0.026$ \citep{dp18},  $>0.016$ \citep{bsf+20}, 0.091 \citep{cma+21} and $>0.079$ \citep{lzc+21}. 
Unlike these, however,  our predictions  are from an optically selected training model 
applied to radio selected samples. Moreover, these include sources with 
up to two missing  magnitudes each.

\section{Conclusions}

The requirement of all nine $FUV,NUV,u,g,r,i,z,W1,W2$ bands  to optimally predict photometric redshifts using
machine learning techniques  can cause a significant reduction in the number of sources to which the model is applicable.
We therefore explore various methods of imputation to replace the missing data, comparing the results 
with those of the non-imputed data. For the SDSS training data, 28\% of the sources have incomplete photometry and
we find simple imputation to be effective, particularly replacing the missing magnitude with the maximum value for that
band. This
method is similar to that of \citet{cma+21}, who assume that missing data are at the detection limit. 
All of the tested methods perform well, however,
giving a regression coefficient of the least-squares linear fit between the predicted and measured redshifts of $r>0.9$ and
a normalised median absolute deviation, which is found to be the best metric to quantify the effect of imputation, 
of $\sigma_{\text{NMAD}} <0.06$.

Our aim is to use the Deep Learning to train models for radio-selected sources, similar to those expected to be detected
with the SKA and its pathfinders. For our three test datasets the number of sources with missing magnitudes ranges from
35\% to 79\%, which would clearly benefit from effective imputation.  Testing these, we find simple imputation to
produce inferior models to multivariate imputation, which uses machine learning to replace the missing values based upon
the features in the data. We suggest that this is since a large fraction of the missing SDSS data will indeed be due to
sensitivity limits (in particular the $u,g,r,i,z$ magnitudes), whereas for the other datasets these may just not have
been observed. This will certainly be the case for the OCARS sources, of which only about half overlap the same region of
sky surveyed by the SDSS with only  half again having all of the magnitudes. This will also be an issue for
the SKA and its pathfinders, although using SkyMapper,  which will observe similar optical bands as the SDSS over the southern sky,
to obtain the photometry will address this.

Testing various levels of imputation, we note a steep decrease in performance when more than two missing magnitudes are
replaced, for all of the test datasets. Also,  the fraction of sources with more than two missing values is small. 
We therefore suggest limiting the imputation to this number per source and, applying this, we find the
performance to be similar to that of the non-imputed data,
although with significant increases in sample sizes. We also find the results  to be comparable to those in the
literature, which all, but one, use only complete data and all of which are tested upon the same dataset used to train
the model.

Of the three training sets, two (FIRST \& LARGESS) do have a large overlap with the SDSS and, with mean radio
flux densities of ${S_{\rm radio}} \sim20$~mJy \citep{cmp21}, these may be representative of 
the radio continuum sources which will be detected with the SKA pathfinders (e.g. \citealt{nha+11}).
The other training set, OCARS, forms a VLBI calibration catalogue and so has much higher fluxes
(${S_{\rm radio}}=340$~mJy). Furthermore, OCARS only overlaps with the SDSS over one
half of the sky. However, if we select OCARS sources with at least one SDSS magnitude, imputing
up to two missing values returns  80\%  of the sample while
yielding an NMAD comparable to those discussed above. The fact that 
application of the Deep Learning method on imputed
data from disparate databases gives similarly good results makes us confident in the applicability of
such techniques to radio sources detected with the SKA and its pathfinders.

\section*{Acknowledgements}

I wish to thank the referee for their prompt and helpful comments.
This research has made use of the NASA/IPAC Extragalactic
Database (NED) which is operated by the Jet Propulsion Laboratory, California Institute of Technology, under contract
with the National Aeronautics and Space Administration and NASA's Astrophysics Data System Bibliographic Service. This
research has also made use of NASA's Astrophysics Data System Bibliographic Service.

\section*{Data availability} 

Data and SDSS {\sf TensorFlow} training model available on request.


\begin{thebibliography}{36}
\expandafter\ifx\csname natexlab\endcsname\relax\def\natexlab#1{#1}\fi

\bibitem[{{Alam} {et~al}\mbox{.}(2015){Alam}, {Albareti}, {Allende Prieto},
  {Anders}, {Anderson}, {Anderton}, {Andrews}, {Armengaud}, {Aubourg},
  {Bailey}, \& et~al.}]{aaa+15}
{Alam} S. {et~al.}, 2015, ApJS, 219, 12

\bibitem[{{Ananna} {et~al}\mbox{.}(2017){Ananna}, {Salvato}, {LaMassa}, {Urry},
  {Cappelluti}, {Cardamone}, {Civano}, {Farrah}, {Gilfanov}, {Glikman},
  {Hamilton}, {Kirkpatrick}, {Lanzuisi}, {Marchesi}, {Merloni}, {Nandra},
  {Natarajan}, {Richards}, \& {Timlin}}]{asu+17}
{Ananna} T.~T. {et~al.}, 2017, ApJ, 850, 66

\bibitem[{{Beck} {et~al}\mbox{.}(2017){Beck}, {Dobos}, {Budav{\'a}ri},
  {Szalay}, \& {Csabai}}]{bdb+17}
{Beck} R., {Dobos} L., {Budav{\'a}ri} T., {Szalay} A.~S., {Csabai} I., 2017,
  Astronomy and Computing, 19, 34

\bibitem[{{Beck} {et~al}\mbox{.}(2021){Beck}, {Szapudi}, {Flewelling},
  {Holmberg}, \& {Magnier}}]{bsf+20}
{Beck} R., {Szapudi} I., {Flewelling} H., {Holmberg} C., {Magnier} E., 2021,
  MNRAS, 500, 1633

\bibitem[{Becker {et~al}\mbox{.}(1995)Becker, White, \& Helfand}]{bwh95}
Becker R.~H., White R.~L., Helfand D.~J., 1995, ApJ, 450, 559

\bibitem[{{Bianchi} {et~al}\mbox{.}(2017){Bianchi}, {Shiao}, \&
  {Thilker}}]{bst17}
{Bianchi} L., {Shiao} B., {Thilker} D., 2017, ApJS, 230, 24

\bibitem[{{Bovy} {et~al}\mbox{.}(2012){Bovy}, {Myers}, {Hennawi}, {Hogg},
  {McMahon}, {Schiminovich}, {Sheldon}, {Brinkmann}, {Schneider}, \&
  {Weaver}}]{bmh+12}
{Bovy} J. {et~al.}, 2012, ApJ, 749, 41

\bibitem[{{Brescia} {et~al}\mbox{.}(2013){Brescia}, {Cavuoti}, {D'Abrusco},
  {Longo}, \& {Mercurio}}]{bcd+13}
{Brescia} M., {Cavuoti} S., {D'Abrusco} R., {Longo} G., {Mercurio} A., 2013,
  ApJ, 772, 140

\bibitem[{{Carvajal} {et~al}\mbox{.}(2021){Carvajal}, {Matute}, {Afonso},
  {Amarantidis}, {Barbosa}, {Cunha}, \& {Humphrey}}]{cma+21}
{Carvajal} R., {Matute} I., {Afonso} J., {Amarantidis} S., {Barbosa} D.,
  {Cunha} P., {Humphrey} A., 2021, A New Window on the Radio Emission from
  Galaxies, Galaxy Clusters and Cosmic Web: Current Status and Perspectives

\bibitem[{{Ching} {et~al}\mbox{.}(2017){Ching}, {Sadler}, {Croom}, {Johnston},
  {Pracy}, {Couch}, {Hopkins}, {Jurek}, \& {Pimbblet}}]{csc+17}
{Ching} J.~H.~Y. {et~al.}, 2017, MNRAS, 464, 1306

\bibitem[{Curran(2020)}]{cur20}
Curran S.~J., 2020, MNRAS, 493, L70

\bibitem[{Curran \& Moss(2019)}]{cm19}
Curran S.~J., Moss J.~P., 2019, A\&A, 629, A56

\bibitem[{Curran {et~al}\mbox{.}(2021)Curran, Moss, \& Perrott}]{cmp21}
Curran S.~J., Moss J.~P., Perrott Y.~C., 2021, MNRAS, 503, 2639

\bibitem[{{D'Isanto} \& {Polsterer}(2018)}]{dp18}
{D'Isanto} A., {Polsterer} K.~L., 2018, A\&A, 609, 111

\bibitem[{{Duncan} {et~al}\mbox{.}(2018){Duncan}, {Brown}, {Williams}, {Best},
  {Buat}, {Burgarella}, {Jarvis}, {Ma{\l}ek}, {Oliver}, {R{\"o}ttgering}, \&
  {Smith}}]{dbw+18}
{Duncan} K.~J. {et~al.}, 2018, MNRAS, 473, 2655

\bibitem[{{Glowacki} {et~al}\mbox{.}(2017){Glowacki}, {Allison}, {Sadler},
  {Moss}, \& {Jarrett}}]{gas+18}
{Glowacki} M., {Allison} J.~R., {Sadler} E.~M., {Moss} V.~A., {Jarrett} T.~H.,
  2017, MNRAS, submitted (arXiv:1709.08634)

\bibitem[{{Han} {et~al}\mbox{.}(2016){Han}, {Ding}, {Zhang}, \&
  {Zhao}}]{hdzz16}
{Han} B., {Ding} H.-P., {Zhang} Y.-X., {Zhao} Y.-H., 2016, Research in
  Astronomy and Astrophysics, 16, 74

\bibitem[{{Hildebrandt} {et~al}\mbox{.}(2010){Hildebrandt}, {Arnouts}, {Capak},
  {Moustakas}, {Wolf}, {Abdalla}, {Assef}, {Banerji}, {Ben{\'\i}tez},
  {Brammer}, {Budav{\'a}ri}, {Carliles}, {Coe}, {Dahlen}, {Feldmann}, {Gerdes},
  {Gillis}, {Ilbert}, {Kotulla}, {Lahav}, {Li}, {Miralles}, {Purger},
  {Schmidt}, \& {Singal}}]{hac+10}
{Hildebrandt} H. {et~al.}, 2010, A\&A, 523, A31

\bibitem[{{Laurino} {et~al}\mbox{.}(2011){Laurino}, {D'Abrusco}, {Longo}, \&
  {Riccio}}]{ldlr11}
{Laurino} O., {D'Abrusco} R., {Longo} G., {Riccio} G., 2011, MNRAS, 418, 2165

\bibitem[{{Li} {et~al}\mbox{.}(2021){Li}, {Zhang}, {Cui}, {Fan}, {Zhao}, {Wu},
  {Zhang}, {Han}, {Xu}, {Tao}, {Li}, \& {He}}]{lzc+21}
{Li} C. {et~al.}, 2021, MNRAS, 509, 2289

\bibitem[{Little \& Rubin(1986)}]{lr86}
Little R. J.~A., Rubin D.~B., 1986, Statistical Analysis with Missing Data.
  John Wiley \& Sons, New York

\bibitem[{{Luken} {et~al}\mbox{.}(2019){Luken}, {Norris}, \& {Park}}]{lnp18}
{Luken} K.~J., {Norris} R.~P., {Park} L.~A.~F., 2019, PASP, 131, 108003

\bibitem[{{Ma} {et~al}\mbox{.}(2009){Ma}, {Arias}, {Bianco}, {Boboltz},
  {Bolotin}, {Charlot}, {Engelhardt}, {Fey}, {Gaume}, {Gontier}, {Heinkelmann},
  {Jacobs}, {Kurdubov}, {Lambert}, {Malkin}, {Nothnagel}, {Petrov},
  {Skurikhina}, {Sokolova}, {Souchay}, {Sovers}, {Tesmer}, {Titov}, {Wang},
  {Zharov}, {Barache}, {Boeckmann}, {Collioud}, {Gipson}, {Gordon}, {Lytvyn},
  {MacMillan}, \& {Ojha}}]{mab+09}
{Ma} C. {et~al.}, 2009, IERS Technical Note, 35, 1

\bibitem[{{Maddox} {et~al}\mbox{.}(2012){Maddox}, {Hewett}, {P{\'e}roux},
  {Nestor}, \& {Wisotzki}}]{mhp+12}
{Maddox} N., {Hewett} P.~C., {P{\'e}roux} C., {Nestor} D.~B., {Wisotzki} L.,
  2012, MNRAS, 424, 2876

\bibitem[{{Malkin}(2018)}]{mal18}
{Malkin} Z., 2018, ApJS, 239, 20

\bibitem[{{Norris} {et~al}\mbox{.}(2011){Norris}, {Hopkins}, {Afonso}, {Brown},
  {Condon}, {Dunne}, {Feain}, {Hollow}, {Jarvis}, {Johnston-Hollitt}, {Lenc},
  {Middelberg}, {Padovani}, {Prandoni}, {Rudnick}, {Seymour}, {Umana},
  {Andernach}, {Alexander}, {Appleton}, {Bacon}, {Banfield}, {Becker}, {Brown},
  {Ciliegi}, {Jackson}, {Eales}, {Edge}, {Gaensler}, {Giovannini}, {Hales},
  {Hancock}, {Huynh}, {Ibar}, {Ivison}, {Kennicutt}, {Kimball}, {Koekemoer},
  {Koribalski}, {L{\'o}pez-S{\'a}nchez}, {Mao}, {Murphy}, {Messias},
  {Pimbblet}, {Raccanelli}, {Randall}, {Reiprich}, {Roseboom},
  {R{\"o}ttgering}, {Saikia}, {Sharp}, {Slee}, {Smail}, {Thompson}, {Urquhart},
  {Wall}, \& {Zhao}}]{nha+11}
{Norris} R.~P. {et~al.}, 2011, PASA, 28, 215

\bibitem[{{P{\^a}ris} {et~al}\mbox{.}(2018){P{\^a}ris}, {Petitjean}, {Aubourg},
  {Myers}, {Streblyanska}, {Lyke}, {Anderson}, {Armengaud}, {Bautista},
  {Blanton}, {Blomqvist}, {Brinkmann}, {Brownstein}, {Brandt}, {Burtin},
  {Dawson}, {de la Torre}, {Georgakakis}, {Gil-Mar{\'\i}n}, {Green}, {Hall},
  {Kneib}, {LaMassa}, {Le Goff}, {MacLeod}, {Mariappan}, {McGreer}, {Merloni},
  {Noterdaeme}, {Palanque-Delabrouille}, {Percival}, {Ross}, {Rossi},
  {Schneider}, {Seo}, {Tojeiro}, {Weaver}, {Weijmans}, {Y{\`e}che}, {Zarrouk},
  \& {Zhao}}]{ppa+18}
{P{\^a}ris} I. {et~al.}, 2018, A\&A, 613, A51

\bibitem[{{Richards} {et~al}\mbox{.}(2001){Richards}, {Weinstein}, {Schneider},
  {Fan}, {Strauss}, {Vanden Berk}, {Annis}, {Burles}, {Laubacher}, {York},
  {Frieman}, {Johnston}, {Scranton}, {Gunn}, {Ivezi{\'c}}, {Nichol},
  {Budav{\'a}ri}, {Csabai}, {Szalay}, {Connolly}, {Szokoly}, {Bahcall},
  {Ben{\'{\i}}tez}, {Brinkmann}, {Brunner}, {Fukugita}, {Hall}, {Hennessy},
  {Knapp}, {Kunszt}, {Lamb}, {Munn}, {Newberg}, \& {Stoughton}}]{rws+01}
{Richards} G.~T. {et~al.}, 2001, AJ, 122, 1151

\bibitem[{{Salvato} {et~al}\mbox{.}(2019){Salvato}, {Ilbert}, \&
  {Hoyle}}]{sih19}
{Salvato} M., {Ilbert} O., {Hoyle} B., 2019, Nature Astronomy, 3, 212

\bibitem[{{Skrutskie} {et~al}\mbox{.}(2006){Skrutskie}, {Cutri}, {Stiening},
  {Weinberg}, {Schneider}, {Carpenter}, {Beichman}, {Capps}, {Chester},
  {Elias}, {Huchra}, {Liebert}, {Lonsdale}, {Monet}, {Price}, {Seitzer},
  {Jarrett}, {Kirkpatrick}, {Gizis}, {Howard}, {Evans}, {Fowler}, {Fullmer},
  {Hurt}, {Light}, {Kopan}, {Marsh}, {McCallon}, {Tam}, {Van Dyk}, \&
  {Wheelock}}]{scs+06}
{Skrutskie} M.~F. {et~al.}, 2006, AJ, 131, 1163

\bibitem[{Troyanskaya {et~al}\mbox{.}(2001)Troyanskaya, Cantor, Sherlock,
  Brown, Hastie, Tibshirani, Botstein, \& Altman}]{tcs+01}
Troyanskaya O., Cantor M., Sherlock G., Brown P., Hastie T., Tibshirani R.,
  Botstein D., Altman R.~B., 2001, Bioinformatics, 17, 520

\bibitem[{van Buuren \& Groothuis-Oudshoorn(2011)}]{vg11}
van Buuren S., Groothuis-Oudshoorn K., 2011, Journal of Statistical Software,
  45, 1

\bibitem[{{Weinstein} {et~al}\mbox{.}(2004){Weinstein}, {Richards},
  {Schneider}, {Younger}, {Strauss}, {Hall}, {Budav{\'a}ri}, {Gunn}, {York}, \&
  {Brinkmann}}]{wrs+04}
{Weinstein} M.~A. {et~al.}, 2004, ApJS, 155, 243

\bibitem[{{White} {et~al}\mbox{.}(1997){White}, {Becker}, {Helfand}, \&
  {Gregg}}]{wbhg97}
{White} R.~L., {Becker} R.~H., {Helfand} D.~J., {Gregg} M.~D., 1997, ApJ, 475,
  479

\bibitem[{{Wolf} {et~al}\mbox{.}(2018){Wolf}, {Onken}, {Luvaul}, {Schmidt},
  {Bessell}, {Chang}, {Da Costa}, {Mackey}, {Martin-Jones}, {Murphy},
  {Preston}, {Scalzo}, {Shao}, {Smillie}, {Tisserand}, {White}, \&
  {Yuan}}]{wol+18}
{Wolf} C. {et~al.}, 2018, PASA, 35, 10

\bibitem[{{Wright} {et~al}\mbox{.}(2010){Wright}, {Eisenhardt}, {Mainzer},
  {Ressler}, {Cutri}, {Jarrett}, {Kirkpatrick}, {Padgett}, {McMillan},
  {Skrutskie}, {Stanford}, {Cohen}, {Walker}, {Mather}, {Leisawitz}, {Gautier},
  {McLean}, {Benford}, {Lonsdale}, {Blain}, {Mendez}, {Irace}, {Duval}, {Liu},
  {Royer}, {Heinrichsen}, {Howard}, {Shannon}, {Kendall}, {Walsh}, {Larsen},
  {Cardon}, {Schick}, {Schwalm}, {Abid}, {Fabinsky}, {Naes}, \&
  {Tsai}}]{wem+10}
{Wright} E.~L. {et~al.}, 2010, AJ, 140, 1868

\end{thebibliography}

\label{lastpage}


\end{document}